\documentclass[aps, reprint,prl,superscriptaddress,nofootinbib]{revtex4-2}
\usepackage{dcolumn}
\usepackage[utf8]{inputenc}
\usepackage{amsmath}
\usepackage{amsfonts}
\usepackage{amssymb}
\usepackage{mathtools}
\usepackage{graphicx}
\usepackage{hyperref}
\usepackage{xcolor}
\usepackage{comment}
\usepackage[toc,page]{appendix}
\usepackage[english]{babel}
\usepackage{amsmath,amssymb,amstext, amsthm}

\newcommand{\ket}[1]{\ensuremath{|{#1}\rangle}}
\newcommand{\bra}[1]{\ensuremath{\langle{#1}|}}

\usepackage{tikz}
\usetikzlibrary{positioning,shapes.geometric, arrows,calc,decorations.text}
\tikzstyle{arrow} = [->,>=stealth]
\tikzstyle{cir}=[inner sep=0,outer sep=0,circle,minimum height=0.1cm,draw=black]
\tikzstyle{blank}=[inner sep=0,outer sep=0,circle,minimum height=0.01cm,draw=black]

\theoremstyle{definition}

\newcommand\expect[1]{\left<#1\right>}

\newcommand\Ev{\mathbb{E}}

\newcommand\ro{{\hat\rho}}
\newcommand\Ho{{\hat H}}
\newcommand\Vo{{\hat V}}
\newcommand\dens{{\varrho}}
\newcommand\denso{{\hat\varrho}}
\newcommand\Phio{{\hat\Phi}}

\newcommand\rv{\mathbf{r}}
\newcommand\sv{\mathbf{s}}
\newcommand\xv{\mathbf{x}}

\newcommand\si{\sigma}

\newcommand\Hcal{\mathcal{H}}

\newcommand{\upd}[0]{\mathrm{d}}

\definecolor{cbl}{rgb}{0,0,1}

\definecolor{crd}{rgb}{1,0,0}

\begin{document}

\title{Collapse-based models for gravity do not violate the entanglement-based witness of non-classicality}

\author{Tianfeng Feng}
 \affiliation{QICI Quantum Information and Computation Initiative,  School of Computing and Data Science, The University of Hong Kong, Pokfulam Road, Hong Kong}

\author{Vlatko Vedral}
\affiliation{Clarendon Laboratory, University of Oxford, Parks Road, Oxford OX1 3PU, United Kingdom}

\author{Chiara Marletto}
\email{chiara.marletto@gmail.com}
\affiliation{Clarendon Laboratory, University of Oxford, Parks Road, Oxford OX1 3PU, United Kingdom}

\date{\today}%

\begin{abstract}
It is known that an entanglement-based witness of non-classicality can be applied to testing quantum effects in gravity. Specifically, if a system can create entanglement between two quantum probes by local means only, then it must be non-classical.
Recently, claims have been made that collapse-based models of classical gravity, i.e., Di\'osi-Penrose model, can predict gravitationally induced entanglement between quantum objects, resulting in gravitationally induced entanglement is insufficient to conclude that gravity is fundamentally quantum, contrary to the witness statement.
Here we vindicate the witness. We analyze the underlying physics of collapse-based models for gravity and show that these models have nonlocal features, violating the assumption of locality. 
We suggest that the entanglement can be generated through quantum-like hidden detectors without interaction with the gravitational field.
\end{abstract}

\maketitle

\section{Introduction}

\begin{figure}
    \centering
    \includegraphics[width=0.95\columnwidth]{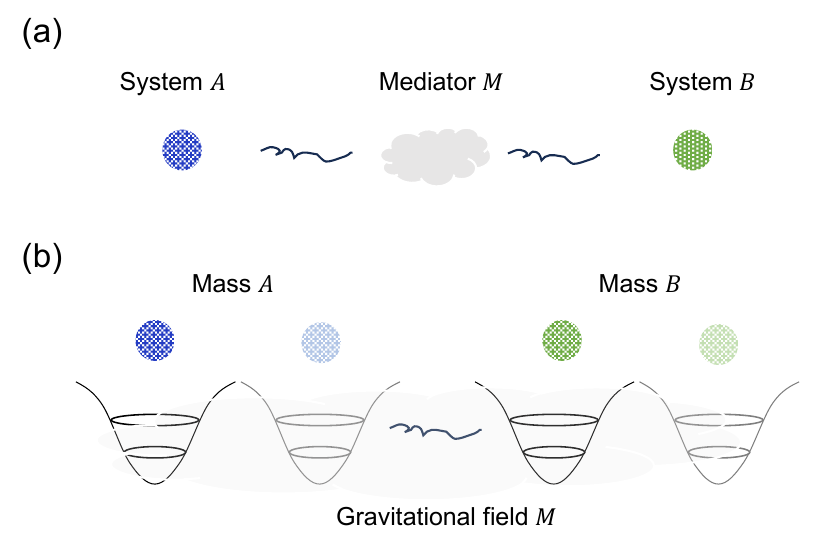}
    \caption{ (a) Schematic representation of the setup for the General Witness Theorem. The two space-like separated quantum probes $A$ and $B$ are coupled only via the unknown mediator $M$, by means of local interactions. Its capability of inducing entanglement between $A$ and $B$ is a witness of its non-classicality. (b) Gravitationally induced entanglement (GIE) between two spatially superposed masses. Observation of GIE suggests the non-classicality of the gravitational field.}
    \label{fig:BMV}
\end{figure}

A particularly promising approach to testing quantum gravity has recently been proposed, based on a novel ``witness of non-classicality''. This witness relies on the entangling power of a given system to conclude that the system has non-classical features. In particular, these tests are based on the so-called General Witness Theorem (GWT), \cite{cmvvRevModPhys.97.015006,marletto_witnessing_2020}, stating that if a system $M$ (such as gravity) can mediate (by local means) entanglement between two quantum systems, $A$ and $B$, (e.g. two masses) then it must be non-classical, \cite{marletto_witnessing_2020}. By ``local means" here we mean a specific protocol, detailed in \cite{marletto_witnessing_2020, bose_spin_2017, marletto_gravitationally_2017}, where $A$ and $B$ must not interact directly with each other, but only via the mediator $M$, as schematically represented in Fig.\ref{fig:BMV} (a). 
Interestingly, ``non-classicality'' is a theory-independent generalisation of what in quantum theory is expressed as ``having at least two distinct physical variables that do not commute", which can be expressed within a general information-theoretic framework, the constructor theory of information, \cite{deutsch_constructor_2015}. Informally, being non-classical means having two or more distinct physical variables that cannot simultaneously be measured to an arbitrarily high degree of accuracy \cite{marletto_witnessing_2020}.

Due to the generality of GWT, it offers a broad theoretical basis for recently proposed experiments that can test quantum effects in gravity at the laboratory scale, based on the generation of gravitational entanglement between two massive probes (see Fig. \ref{fig:BMV} (b)) - the so-called Bose-Marletto-Vedral effect, \cite{bose_spin_2017,marletto_gravitationally_2017}. It also provides a basis for any other experiment that (beyond the case of gravity) intends to show that some system $M$ is non-classical \cite{raia_role_2024}, using the effect of its entangling power. 

A particularly appealing feature of the GWT is that, by using the constructor theory of information, it avoids assuming the usual machinery of quantum information theory, thus extending beyond quantum theory existing results, such as the theorems that forbid the creation of entanglement via local operations and classical communication. Moreover, the GWT is proven without assuming the existence of a probability space, in contrast to existing approaches such as Generalised Probabilistic Theories \cite{plavala_general_2023}. This generality is particularly important as one wants to use it in a context where the system $M$ may or may not obey quantum theory itself. 

This witness relies on the capacity of a system $M$ to generate entanglement between two independent subsystems, initially unentangled with each other. For the witness to be applicable, it is key that the systems $A$ and $B$ are independent -- hence it is essential to assume the principle of locality.\\

Recently, some claims have been made that the collapse-based model of classical gravity \cite{penrose1996gravity,diosi1987universal,DiosiPhysRevA.40.1165,Tilloy_2016}, i.e. Di\'osi-Penrose model, can predict gravitationally induced entanglement between quantum objects \cite{trillo2024diosipenrosemodelclassicalgravity, angeli2025positivityentanglementmarkovianopen, Liu_2023, LiuPhysRevD.111.062004}, resulting in gravitationally induced entanglement is insufficient to conclude that gravity is fundamentally quantum \cite{trillo2024diosipenrosemodelclassicalgravity}. 
These claims are contrary to the GWT statement. 
Here we vindicate the GWT and 
show that collapse-based models of gravity are fundamentally nonlocal, violating the assumption of locality.
The generation of entanglement between masses stems from the hidden quantum-like monitoring process, not from the backaction of the gravitational field. That is, the gravitational field does not induce the entanglement.
Even though the interaction between quantized matter and the gravitational field cannot generate entanglement, the backaction of the gravitational field is nonlocal itself. Besides, the monitoring process as a part of the dynamics violates the principle of locality and reveals quantum-like features.

In other words, collapse-based models of gravity do not violate the GWT for non-classicality.  Since classical theories usually lack nonlocal and entanglement properties, it is worth questioning whether the mechanism of collapse-based models for gravity, i.e., the DP model, should still be classified as generically classical for gravity.

\section{Brief Review of collapse-based models of gravity}

Here we briefly review the collapse-based models of gravity \cite{diosi1987universal,DiosiPhysRevA.40.1165, penrose1996gravity,Bahrami_2014}. Through the lens of continuous measurement theory, Tilloy and Di\'osi present a linear version of collapse-based models, i.e. DP model
\cite{Tilloy_2016}.

The collapse-based models are equivalent to the continuous monitoring of the mass density by hidden detectors of finite spatial resolution $\sigma$.
In this model, the detectors can be entangled, correlating their measurement outcomes. Tilloy and Di\'osi refer to this continuous equivalent of a von Neumann measurement result as the ``signal,'' which can subsequently be used to feedback a back-action on the gravitational field when Continuous spontaneous localization (CSL) and DP models are considered. The signal in continuous measurement theory is defined as follows:
\begin{equation}
    \dens_t(\rv)=\langle\denso_\si\rangle_t +\delta\dens_t(\rv),
    \label{signal}
\end{equation}
where 
$\hat{\dens}_\sigma(\rv)=\sum_{n=1}^N m_n  g_\si(\rv-\hat\xv_n)$ ($N$ is the particle number) representing density of matter
and  
$\delta\dens_t(\rv)$ denotes the white noise in time, satisfying $
\Ev[\delta\dens_t(\rv) \delta\dens_\tau(\sv)]=\gamma^{-1}_{\rv\sv}\delta(t-\tau)
$. Note that the function $g_\sigma$
is a normalized Gaussian of width $\sigma$ \cite{Tilloy_2016}.
Here, $\gamma_{rs}$ is a non-negative kernel that intuitively encodes the correlation between the detectors at positions $r$ and $s$.
After monitoring by a hidden detector, the stochastic master equation (SME) that governs the dynamics of the density matrix of the system is given by \cite{Tilloy_2016}:
\begin{eqnarray}\label{SME0}
\frac{\upd\ro_{SEM}}{\upd t}&=&-i[\Ho,\ro]-\int \upd\rv\upd\sv\frac{\gamma_{\rv\sv}}{8}[\denso_\si(\rv),[\denso_\si(\sv),\ro]]\nonumber\\
                            & & +\int \upd\rv\upd\sv\frac{\gamma_{\rv\sv}}{2}\Hcal[\denso_\si(\rv)]\ro \delta\dens(\sv),
\end{eqnarray}
where $\Hcal[\denso_\si(\rv)](\hat\rho)=\left\{\denso_\si(\rv)-\expect{\denso_\si(\rv)}_t,\ro_t \right\}$ and $(\si)$ denotes an optional convolution with $g_\si$.
This is the standard SME if one replaces $\int dr \ro_{\sigma}(r)$ as another operator $\hat{A}$ \cite{wiseman2009quantum}. Note that we have set $\hbar=1$.
If we average the measurement outcome of the hidden detector, the state
satisfies

\begin{eqnarray}\label{SME}
\frac{\upd\ro_{SEM}}{\upd t}&=&-i[\Ho,\ro]-\int \upd\rv\upd\sv\frac{\gamma_{\rv\sv}}{8}[\denso_\si(\rv),[\denso_\si(\sv),\ro]]\nonumber.\\
\end{eqnarray}
Now, the provided equation is defined in the absence of gravitational effects. 
This constitutes the monitoring process of the density of quantum matter.

For considering the gravitational effects, the back-action of the quantum matter on the classical gravitational field should be analysed. Compared to directly making use of the Poisson equation $\nabla^2 \Phi(\rv)=4\pi G {\expect{\denso(\rv)}}$ to source the classical Newton potential (we don't consider GR effect), one can utilize Eq. (\ref{signal}) instead of $\expect{\denso(\rv)}$, e.g.
 \begin{equation}
 \nabla^2 \Phi(\rv)=4\pi G {\dens(\rv)}.
 \end{equation}
The modified semiclassical Newton potential now becomes stochastic and takes the form 
$\Phi(\rv)=-G\int\upd\sv\frac{\dens(\sv)}{\vert\rv-\sv\vert}$ \cite{Tilloy_2016}.
That is, the semiclassical interaction $\Vo_G$ is now given as
 \begin{equation}\label{V}
\Vo_G=\int\upd\rv\Phi(\rv)\denso_{(\si)}(\rv)=\int\upd\rv\dens(\rv)\Phio_{(\si)}(\rv).
 \end{equation}
At this moment, the quantum matter will evolve according to 
 $e^{-i\Vo_G dt}$. 
Thus, the dynamics of collapse-based models include two processes:

\textbf{(a)} The first one is the \textbf{monitoring process} and the evolution of quantum matter, which satisfies SME (Eq. \ref{SME}). 

\textbf{(b)} The second one concerns the \textbf{back-action} on the gravitational field and the sequential evolution, for instance, $e^{-i\Vo_G dt}$ acting on the evolved state $\ro+d \ro^{SME}$. 

Note that there is a temporal order for these two processes.
Specifically, a short-time evolution state of the quantum matter is 
\begin{equation}\label{final}
\ro+\upd\ro=e^{-i\Vo_{G} \upd t}(\ro+\upd\ro^{\text{SME}})e^{i\Vo_{G} \upd t},
\end{equation}
where $\upd\ro^{\text{SEM}}$ is the change of quantum state according to Eq. (\ref{SME}) in the monitoring procese.
Expanding the exponential up to second order in Eq. (\ref{final}) then gives the average SEM for the complete evolution\cite{Tilloy_2016}:

\begin{eqnarray}\label{SMEfull}
&&\frac{\upd\ro_{final}}{\upd t}=-i\left[\Ho+\Vo_{G,\si},\ro\right]\nonumber\\
&& \quad \quad\quad \hskip-24pt-\!\!\!\int \!\!\!\upd\rv\upd\sv\!\left(\!\!\frac{\gamma_{\rv\sv}}{8}[\denso_\si(\rv),[\denso_\si(\sv),\ro]]
                                          \!+\!\frac{\gamma_{\rv\sv}^{-1}}{2}[\Phio_{(\si)}(\rv),[\Phio_{(\si)}(\sv),\ro]]\!\!\right)\nonumber,\\
\end{eqnarray}
where 
$\Vo_{G,\si}=-\frac{G}{2}\int \upd\rv\upd\sv \frac{\denso_\si(\rv)\denso_{(\si)}(\sv)}{\vert\rv-\sv\vert}
$ is an effective potential term.
The CSL and DP models both adhere to the stochastic master equation of Eq. (\ref{SMEfull}), with the primary distinction between them being the choice of the nontrivial correlator  $\gamma_{\rv\sv}$\cite{Tilloy_2016}. For the CSL model, the correlator is chosen as $\gamma_{\rv\sv}=\gamma \delta(\rv-\sv)$. For DP model, $\gamma_{\rv\sv}=\kappa G\frac{1}{\vert\rv-\sv\vert}$ or $\gamma_{\rv\sv}^{-1}=-\frac{1}{4\pi\kappa G}\delta(t-\tau)\nabla^2\delta(\rv-\sv)$.

\section{Entanglement generation in the DP model}

For analysing the dynamics of the DP model in detail, we set $\gamma_{\rv\sv}=2 G\frac{1}{\vert\rv-\sv\vert}$ and substitute it into Eq. (\ref{SMEfull}), resulting evolution of the equation of the DP model,

\begin{equation}
\frac{\upd\ro_{final_{DP}}}{\upd t}=-i\left[\Ho,\ro\right]\nonumber + \frac{G}{2}A(\ro), \\
\label{SMEfullDP}
\end{equation}
where 
\begin{equation}
\begin{split}
       A(\rho)&= -\int d\rv d\sv \frac{1}{|\rv-\sv|}
[\denso_\si(\rv),[\denso_\si(\sv),\ro]]\\
&+i \int d\rv d\sv \frac{1}{|\rv-\sv|} [\denso_\si(\rv)\denso_\si(\sv),\ro].
\end{split}
\end{equation}
This equation has the same form as in \cite{trillo2024diosipenrosemodelclassicalgravity}.

Consider two particles (particle 1 and particle 2) with mass $m$, each in a superposition of two positions, similar to the spatial superposition in the BMV experiment (see figure \ref{fig:BMV} (b)). The position of particle 1 is $\mathbf{r}_1 = \pm \frac{a}{2} $ (spacing $a$), and the position of particle 2 is $\mathbf{r}_2 = \pm \frac{a}{2}  + d $ (center-to-center distance $d$). The initial state is a product superposition of the two particles:
\begin{equation}
    |\psi(0)\rangle = \frac{1}{2} \left( |0\rangle_1 + |1\rangle_1 \right) \otimes \left( |0\rangle_2 + |1\rangle_2 \right),
    \label{rho0}
\end{equation}
with the corresponding density matrix:
$ \rho(0) = \frac{1}{4} \sum_{x_1,x_2} |x_1\rangle\langle x_1| \otimes |x_2\rangle\langle x_2|, $
where $x_i \in \{0, 1\}$.
In \cite{trillo2024diosipenrosemodelclassicalgravity,angeli2025positivityentanglementmarkovianopen}, the authors show that the collapsed-based models of gravity, i.e. DP model, can predict entanglement between masses 1 and 2, which may conflict with GWT.

In the following analysis, we will show how the DP model processes nonlocal features and violates the principle of locality. Thus, the collapsed-based models do not violate the GWT. 
 We emphasize that, for analysing the underlying physics of the collapse-based models of gravity, Eqs. (\ref{SME}) and (\ref{final}) are fundamental. This is because Eq. (\ref{SMEfull}) and Eq. (\ref{SMEfullDP}) are derived from these equations.

\section{What goes wrong in the argument}

Recently, Trillo and Navascu\'e claim the gravitationally induced entanglement is insufficient to conclude that gravity is fundamentally quantum since the DP model can generate gravitationally induced entanglement between two quantum masses \cite{trillo2024diosipenrosemodelclassicalgravity}.
Here we disprove their claims and show that their work does not violate the entanglement-based witness of non-classicality, e.g., GWT. In the DP model \cite{Tilloy_2016,trillo2024diosipenrosemodelclassicalgravity}, the generation of entanglement between two masses is not induced by classical gravity but instead by a hidden monitoring process with quantum features. The DP model is fundamentally nonlocal or has a quantum degree of freedom to mediate the interaction between matter.
Thus, the claim that ``\emph{the Di\'osi-Penrose model of classical gravity predicts gravitationally induced entanglement}'' is inaccurate and not true. 

Here we refute the argument in question. 

\textbf{1. About the gravitational field.} The dynamics of collapse-based models of gravity include two processes.
The first one is the monitoring process (which can weakly measure the density) and the evolution of quantum matter satisfying the SME Eq. (\ref{SME}).  The second one concerns the back-action on the gravitational field and the sequential evolution. The interaction $e^{-i\Vo_Gt}$ induced by a classical gravitational field is a local unitary (see Eq. (\ref{V})), and it can not generate entanglement between quantum matters. Since $\hat{\dens}_\sigma(\rv)=\sum_{n=1}^N m_n  g_\si(\rv-\hat\xv_n)$, and $[g_\si(\rv-\hat\xv_i),g_\si(\rv-\hat\xv_{j})]=0$ with $i\ne j$,   
\begin{equation}
    e^{-i\Vo_Gt}=\prod_{n=1}^{N} e^{-i\int\upd\rv\Phi(\rv)m_n g_\si(\rv-\hat\xv_n)}.
\end{equation}
On the other hand, from the quantum information perspective, if the gravitational field is classical (scalar function $\Phi(\rv)$ and it must have only one basis $\{\ket{i}\bra{i}\}$). As an entity, the purely classical gravitational field can not mediate entanglement between two quantum masses \cite{marletto_gravitationally_2017,cmvvRevModPhys.97.015006}. One more point needs to be clarified: even though the back-action of the gravitational field in the DP model can not generate entanglement, it is fundamentally nonlocal.\\

\textbf{2. About the entanglement generation.}
The entanglement between masses can be generated in the DP model \cite{trillo2024diosipenrosemodelclassicalgravity,angeli2025positivityentanglementmarkovianopen}, but it is not induced by gravity.
 The nonrival hidden detector induces the entanglement of the DP model \cite{Tilloy_2016,trillo2024diosipenrosemodelclassicalgravity,angeli2025positivityentanglementmarkovianopen} with the non-trivial choice of $\gamma_{\rv\sv}=2 G\frac{1}{\vert\rv-\sv\vert}$. This correlator is nonlocal, as mentioned in other nonlinear gravity works \cite{Bahrami_2014}.

 We reveal that the monitoring process by hidden detectors in the DP model, as described by the SME (Eq. \ref{SME}), can generate entanglement. This master equation is fundamentally nonlocal since the ``jump'' term carries a nonlocal correlator $\frac{1}{|\rv-\sv|}$. An intuitive explanation is that the 'jump' term in the master equation can connect emissions from two distant locations. Specifically, if an emission is superposed across two locations and we detect it, this detection will induce a backaction on the systems at those locations, resulting in their entanglement.
 
 Without loss of generality, set $H=0$, and
expand to first order in \( \Delta t \), ignoring higher-order terms, one has
\begin{equation}
    \begin{split}
        \ro(\Delta t) &\approx \ro(0) + \frac{d}{dt}\ro_{SME}\Delta t\\
       & \approx \ro(0) -\int \upd\rv\upd\sv\frac{G}{4|\rv-\sv|}[\denso_\si(\rv),[\denso_\si(\sv),\ro]]\Delta t .
    \end{split}
\end{equation}

Our goal here is to analyze the degree of entanglement of $\ro(\Delta t)$.
Suppose there is a two-particle system with initial state $\ro(0)=\ket{\psi(0)}\bra{\psi(0)}$, where 
    $|\psi(0)\rangle = \frac{1}{2} \left( |0\rangle_1 + |1\rangle_1 \right) \otimes \left( |0\rangle_2 + |1\rangle_2 \right) $.
Here, the mass density is given as,
\[
\denso_\sigma (\rv) = \denso_1(\mathbf{r}) + \denso_2(\mathbf{r}) \\
=\sum_{n=1}^2 m_n  g_\si(\rv-\hat\xv_n),
\]
where $m_1=m_2$ and $g_\sigma(\rv-\xv_i)=\frac{e^{-|\mathbf{r} - \mathbf{x}_i|^2/(2\sigma)}}{(2\pi \sigma)^{3/2}}$. Note that $I_{\xv_i,\xv_j}=\int dr ds \frac{1}{|{\rv-\sv}|}g_{\sigma}(\rv-\xv_i)g_{\sigma}(\sv-\xv_j)=\frac{\text{erf}(|\xv_i-\xv_j|/(2\sigma))}{|\xv_i-\xv_j|} \quad \text{or} \quad \frac{1}{\sigma\sqrt{\pi}}$ for $|\xv_i-\xv_j|\ne 0$ and $|\xv_i-\xv_j|= 0$ respectively \cite{trillo2024diosipenrosemodelclassicalgravity}, where $\text{erf}(x)$ is the error function.
Without loss of generality, one may set $\Tilde{f}(z)=\text{erf}(z/2\sigma)/z$ if $z\ne 0$ and $\Tilde{f}(z)= \frac{1}{\sigma\sqrt{\pi}}$ if $z=0$.
As shown in Figure \ref{fig:BMV} (b), in the Bose-Marletto-Vedral (BMV) experiments, two particles are on the same horizontal line, so we can replace $\xv$ with the scalar $x$.
One can directly compute the density matrix element as: 

\begin{equation}
\begin{split}
     &\quad \bra{x_1,x_2}\ro(\Delta t)\ket{y_1,y_2}\\
     &= [1
     -\frac{Gm^2 \Delta t}{4}(\sum_{i,j=1}^2 I_{x_i,x_j}+I_{y_i,y_j}-2I_{x_i,y_j})]\\
     & \quad \cdot \bra{x_1,x_2}\ro(0)\ket{y_1,y_2},
\end{split}
\end{equation}
where $x_{1}, x_2,y_{1},y_2\in \{0,1\}$ representing the position of the corresponding particle. Since  $\mathbf{x}_1 = \pm \frac{a}{2} $ ($\ket{0}_1:=\ket{x_1=-\frac{a}{2}}$ and $\ket{1}_1:=\ket{x_1=+\frac{a}{2}}$), and  $\mathbf{x}_2 = \pm \frac{a}{2}  + d $ ($\ket{0}_2:=\ket{x_2=d-\frac{a}{2}}$ and $\ket{1}_2:=\ket{x_2=d+\frac{a}{2}}$),
after performing the partial transpose on particle~2 and diagonalizing the first-order correction in the zero eigenspace, the negativity of $\rho(\Delta t)$ is found to be
\begin{equation}
    \begin{split}
       \mathcal{N}(\rho^{T_2}(\Delta t)) &= \frac{G m^2 \Delta t}{4\hbar}\Biggl[\bigl(\tilde{f}(0)-\tilde{f}(a)\bigr) +  \\& \frac12\Bigl(\tilde{f}(d-a)+\tilde{f}(d+a)-2\tilde{f}(d)\Bigr)\Biggr] ,
    \end{split}
\end{equation}
where $\tilde{f}(z) = \text{erf}(z/(2\sigma))/z$ and $\tilde{f}(0)=1/(\sigma\sqrt{\pi})$. This total expression is manifestly non‑negative for all parameters. If we shift the coordinate of particle~2 by $a$, i.e., let $d \to d+a$, then the expression reduces to the one reported in Ref.\cite{trillo2024diosipenrosemodelclassicalgravity}, with distances $d$, $d+a$, and $d+2a$. In the limit $d\to\infty$ the second bracket vanishes, leaving a constant positive term proportional to $\tilde{f}(0)-\tilde{f}(a) >0$, which reflects the long‑range spatial correlation of the hidden detectors in the DP model. Hence, entanglement is already generated at first order in $\Delta t$ by the monitoring process alone.



\textbf{3. About the quantum hidden degree of freedom in the monitoring process.} In the collapse-based model of gravity \cite{trillo2024diosipenrosemodelclassicalgravity,Tilloy_2016,angeli2025positivityentanglementmarkovianopen}, the SEM (see Eq. \ref{SME}) is used to analyze the DP and CML models. This SME stems from the continuous measurement theory in which there is a hidden degree of freedom that can monitor the observable of interest \cite{Tilloy_2016}. In the DP case, this hidden degree of freedom (also called `hidden detector') weakly measures the mass density. 
It is well known that there is no consistent quantum-classical dynamics for quantum measurement problems \cite{diosi2000quantum,diosi2014hybrid}. Solving the measurement problem requires either introducing non-commutative degrees of freedom into classical degrees of freedom (which contradicts the definition of classical systems) or the introduction of superselection rules \cite{sudarshan1976interaction,koopman1931hamiltonian}. 
Indeed, continuous measurement theory is also introduced to address the quantum measurement problem, and it introduces a quantum hidden detector \cite{wiseman2009quantum}.
Thus, this hidden degree of freedom must be quantum-like. 
If the hidden degree of freedom is classical, it can not be used to monitor the observables of quantum systems. 
In \cite{Tilloy_2016}, the author made the following statement:\\

\emph{``We now consider a general many-particle spontaneous localization model which includes CSL and DP as specific cases. Formally, it is equivalent to the continuous monitoring of the mass density by (hidden) detectors of spatial resolution $\sigma$. The detectors are also possibly entangled, which correlates with their measurement outcomes.''}\\

What does this hidden degree of freedom belong to in a composite system comprising the masses and the gravitational field? One possibility is that it should be part of the gravitational field, as it mediates quantum information between all the masses. In this scenario, the gravitational field possesses a hidden quantum degree of freedom, making it a truly quantum object.

Conversely, if the quantum degree of freedom does not reside within the gravitational field, it must be distributed throughout space to effectively capture matter-related information. This hidden degree of freedom should also uphold internal locality principles, facilitating quantum information exchange between masses via the classical field. Otherwise, it would violate the principle of locality.

If the monitoring process of hidden detectors is not a physical process but merely a mathematical construct \cite{Tilloy_2016, trillo2024diosipenrosemodelclassicalgravity,angeli2025positivityentanglementmarkovianopen}, collapse-based models would violate causality and the principle of locality, as they implicitly assume instantaneous interactions between masses. This assumption further leads to the paradoxical conclusion that all matter would spontaneously decohere and (weakly) entangle together.

In summary, the DP model appears to maintain the classical nature of the gravitational field, but its dynamics implicitly achieve quantization through globally correlated signals, violating the principle of locality. Therefore, although the DP model claims to describe ``classical gravity'', its actual dynamics implicitly incorporate features of the quantum nature of gravity.

\section{Discussions}

In \cite{Diosi_PhysRevA.107.062206, trillo2024diosipenrosemodelclassicalgravity, Tilloy_2016, PhysRevX.13.041040,angeli2025positivityentanglementmarkovianopen}, the authors utilize a general master equation (which satisfies the completely positive map and maintains the evolution state as a classical-quantum state) to investigate the interaction between gravity and quantum matter. These equations suggest that there is a hidden quantum degree of freedom participating in the interaction process between gravity and quantum matter. While some may argue that this hidden degree of freedom is a mathematical construct rather than a physical entity, this argument is unconvincing and presents conceptual and potential causal issues.

Focusing solely on the master equation of quantum subsystems or adding correction terms only to these equations can obscure the non-classical behavior of the gravitational field or introduce a hidden degree of freedom (the third entity). Therefore, the dynamic analysis of embedding classical systems into quantum systems within a unified context is essential. Furthermore, these modified master equations should ideally be describable by a Schrödinger equation for a larger closed system, where the Hamiltonian of this larger system still comprises matter, the (extended) gravitational field, and their interaction terms. In this case, if there is a backaction from the quantum matter to the gravitational field, this implies the quantum-like nature of gravity \cite{feng2024conservationlawsquantizationgravity}.

Collapse-based models of gravity \cite{diosi1987universal,DiosiPhysRevA.40.1165,trillo2024diosipenrosemodelclassicalgravity, Tilloy_2016,angeli2025positivityentanglementmarkovianopen} fall into the category of models that are not fundamentally semiclassical. These models introduce a hidden quantum degree of freedom in the coupling of quantum objects and the gravitational field. Our analysis and conclusions may also apply to other stochastic gravity models \cite{PhysRevX.13.041040}.

From the perspective of both theoretical predictions and experiments, true gravitationally induced entanglement can be readily distinguished from the entanglement generated by the DP model \cite{angeli2025positivityentanglementmarkovianopen}. Consequently, the entanglement-based tests of gravity can falsify the DP model and non-linear models entirely \cite{angeli2025positivityentanglementmarkovianopen,trillo2024diosipenrosemodelclassicalgravity,Liu_2023,LiuPhysRevD.111.062004}.

\section{Conclusions}

With the assumption of the principle of locality, gravitationally induced entanglement is the sufficient condition for the quantum nature of gravity \cite{marletto_gravitationally_2017,bose_spin_2017,cmvvRevModPhys.97.015006}, i.e., the GWT holds \cite{cmvvRevModPhys.97.015006}. 
Collapse-based models of gravity, such as the DP model, do not invalidate the GWT since they somehow violate the principle of locality or introduce a hidden quantum degree of freedom to mediate the quantum information of quantum matter.

The DP model, typically viewed as a classical model of gravity, incorporates quantum degrees of freedom during its construction, potentially leading to nonlocal effects or entanglement. This departure from classical behavior raises the question of whether it can still be classified as a purely classical gravitational model. \\

\textbf{Note added.--}
Recently, Aziz and Howl \cite{aziz2025classical} claimed, using a quantum field–theoretic approach, that classical gravity can generate entanglement via local interactions, a claim that has prompted several refutations \cite{marletto2025classicalgravity1,diósi2025noclassicalgravitydoes,marletto2025classicalgravity2}. Our work analyzes collapse‑based models of classical gravity and shows that, in such models, entanglement arises from a nonlocal map that cannot result from purely local gravitational interactions. This analysis supports the validity of the BMV experiment as a test of the quantum nature of gravity.

{\bf Acknowledgements.--} 
We thank Di\'osi Lajos, Alan Forrester, and Matteo Carlesso for their comments on the manuscript. T. Feng thanks Qiuyi Ma, Qian Chen, and Yubao Liu for their helpful discussion. This research was made possible through the generous support of the Gordon and Betty Moore Foundation, the Eutopia Foundation, and of the ID 62312 grant from the John Templeton Foundation, as part of the \href{https://www.templeton.org/grant/the-quantuminformation-structure-ofspacetime-qiss-second-phase}{‘The Quantum Information Structure of Spacetime’ Project (QISS)}. The opinions expressed in this project/publication are those of the author(s) and do not necessarily reflect the views of the John Templeton Foundation. 

\bibliographystyle{apsrev4-2}
\bibliography{locality}

\end{document}